\documentclass[fleqn,11pt]{wlscirep}
\usepackage[T1]{fontenc}
\usepackage{hyperref}
\usepackage{bm} 
\usepackage{amsmath}
\newcommand{\nc}{\newcommand*} 
\usepackage{hyperref}  
\usepackage{cleveref}  

\usepackage{changes}

\usepackage{tikz}
\usepackage{graphicx}

\nc{\figurewidth}{3.2in}
\nc{\xbar}{\bar{x}}
\nc{\rhoeq}{\rho_{\mathrm{eq}}}
\nc{\zeq}{z_{\mathrm{eq}}}
\nc{\tla}{\tilde{\lambda}}
\nc{\bt}{\beta}
\nc{\dt}{\delta}
\nc{\Dt}{\Delta}
\nc{\vj}{\vec{j}}
\nc{\vl}{\vec{l}}
\nc{\hx}{\hat{x}}
\nc{\hy}{\hat{y}}
\nc{\bj}{\bm{j}}
\nc{\mJ}{\mathcal{J}}
\nc{\mP}{\mathcal{P}}
\nc{\Msun}{M_\odot}
\nc{\app}{\approx}
\nc{\av}[1]{\langle #1 \rangle}
\nc{\eq}[1]{Eq.~\eqref{#1}}
\nc{\al}{\alpha}
\nc{\Xstar}{X_{\ast}}
\nc{\fpbh}{f_{\mathrm{pbh}}}
\nc{\vth}{\vec{\theta}}
\nc{\vla}{\vec{\lambda}}
\nc{\vd}{\vec{d}}
\nc{\Mmin}{M_{\mathrm{min}}}
\nc{\rmd}{\mathrm{d}}
\nc{\mmin}{{m_{\mathrm{min}}}}
\nc{\mmax}{{m_{\mathrm{max}}}}
\nc{\mR}{\mathcal{R}}
\nc{\tmR}{\tilde{\mathcal{R}}}
\nc{\s}{\sigma}
\nc{\ogw}{\Omega_{\mathrm{GW}}}
\nc{\addref}{[\textcolor{red}{add ref}] }
\nc{\Om}{\Omega}
\nc{\eps}{\epsilon}
\nc{\gm}{\gamma}
\nc{\Gm}{\Gamma}
\nc{\gpcyr}{\mathrm{Gpc}^{-3}\,\mathrm{yr}^{-1}}
\nc{\Eq}[1]{Eq.~\eqref{#1}}
\nc{\Fig}[1]{Fig.~\ref{#1}}
\nc{\Table}[1]{Table~\ref{#1}}
\nc{\lvc}{LIGO/Virgo} 
\nc{\Sec}[1]{Sec.~\ref{#1}}
\nc{\eg}{\textit{e.g.~}}
\nc{\SNR}{\mathrm{SNR}}
\nc{\be}{\bm{\epsilon}}
\nc{\mb}{\bm}
\nc{\bbt}{\bm{t}}
\nc{\bth}{\bm{\theta}}
\nc{\bn}{\bm{n}}
\nc{\bd}{\bm{d}}
\nc{\ba}{\bm{a}}
\nc{\bep}{\bm{\epsilon}}
\nc{\bnu}{\bm{\nu}}
\nc{\uni}{\mathrm{U}}
\nc{\logu}{\operatorname{\mathrm{log-U}}}
\nc{\RN}{\mathrm{RN}}
\nc{\BN}{\mathrm{BN}}
\nc{\GN}{\mathrm{GN}}
\nc{\mcN}{\mathcal{N}}
\nc{\GWB}{\mathrm{GW}}
\nc{\yr}{\mathrm{yr}}
\nc{\Am}{\mathcal{A}}
\nc{\Dm}{\mathcal{D}}
\nc{\Hm}{\mathcal{H}}
\nc{\sovast}{Soviet Ast.}
\nc{\apj}{Astrophys. J.}
\nc{\ISCb}{ISC\textsubscript{b}}
\nc{\Monond}{Monopole\textsubscript{2nd}}

\nc{\mrm}{\mathrm}
\nc{\BE}{B\scriptsize{AYES}\normalsize{E}\scriptsize{PHEM}\normalsize  }

\nc{\Ostgw}{\Omega_{\mathrm{GW}}^{\mathrm{ST}}}
\nc{\Ottgw}{\Omega_{\mathrm{GW}}^{\mathrm{TT}}}
\nc{\Ovlgw}{\Omega_{\mathrm{GW}}^{\mathrm{VL}}}
\nc{\Oslgw}{\Omega_{\mathrm{GW}}^{\mathrm{SL}}}

\def\({\left(}
\def\){\right)}
\def\[{\left[}
\def\]{\right]}

\def\e{\begin{equation}}
	\def\q{\end{equation}}
\def\m{\begin{eqnarray}}
	\def\n{\end{eqnarray}}
\nc{\red}[1]{\textcolor{red}{#1}}

\title{The spatial correlations between pulsars for interfering sources in Pulsar Timing Array and evidence for gravitational-wave background in NANOGrav 15-year data set}

\author[1,2,3,*]{Yu-Mei Wu}
\author[3,4,*]{Yan-Chen Bi}
\author[2,3,4,$\dagger$]{Qing-Guo Huang}
\affil[1]{Center for Gravitation and Cosmology,
College of Physical Science and Technology,
Yangzhou University, Yangzhou, 225009, China}
\affil[2]{School of Fundamental Physics and Mathematical Sciences,
    Hangzhou Institute for Advanced Study, UCAS, Hangzhou 310024, China}
\affil[3]{School of Physical Sciences, University of Chinese Academy of Sciences, No. 19A Yuquan Road, Beijing 100049, China}
\affil[4]{CAS Key Laboratory of Theoretical Physics, Institute of Theoretical Physics, Chinese Academy of Sciences, Beijing 100190, China}
\affil[*]{These authors contributed equally to this work: Yu-Mei Wu and Yan-Chen Bi}
\affil[$\dagger$]{Corresponding author: huangqg@itp.ac.cn}

\begin{abstract}
Pulsar timing arrays (PTAs), aimed at detecting gravitational waves (GWs) in the $1\sim 100$ nHz range, have recently made significant strides. Compelling evidence has emerged for a common spectrum signal spatially correlated among pulsars, following a Hellings-Downs (HD) pattern, which is crucial for detecting a gravitational-wave background (GWB). However, the HD curve is expected for discrete and non-interfering sources, which is unlikely to hold in realistic scenarios with potential interference among numerous GW sources, such as the supermassive black-hole binaries. Incorporating interference was previously expected to introduce an irreducible uncertainty (known as "cosmic variance") in discerning the HD correlation; however, our work reveals how this interference generates measurable frequency-dependent spatial correlations distinct from the HD curve. The spatial correlations for interfering sources (referred to as ``ISC") still exhibit contributions in the quadrupole and higher orders, resembling the HD correlation and encoding the nature of GW radiations. We apply these novel correlations to search for a GWB in the NANOGrav 15-year data set. In an optimistic estimation, our findings show a Bayes factor of $33.7\pm 3.2$ comparing ISC to the HD correlation, and an improvement in optimal statistic signal-to-noise ratio from $4.9\pm 1.1$ for the HD correlation to $6.6\pm 1.7$ for the ISC, highlighting the significant enhancement in evidence for detecting a GWB through incorporating interference.
\end{abstract}
\begin{document}

\flushbottom
\maketitle

\thispagestyle{empty}

\noindent Pulsar Timing Arrays (PTAs)~\cite{1990ApJ...361..300F} serve as expansive gravitational-wave (GW) detectors on a galactic scale. These arrays are dedicated to the detection of low-frequency GWs, typically in the nanohertz range, through precise monitoring of pulse arrival times from a constellation of pulsars~\cite{Detweiler:1979wn,1978SaPTA,Burke-Spolaor:2018bvk}. The influence of GWs, such as those emitted by inspiraling supermassive black-hole binaries at the cores of merging galaxies, manifests in the timing of these pulses in a specific correlated manner across multiple pulsars~\cite{Begelman:1980vb,Mingarelli:2017fbe,NANOGrav:2023hfp,Bi:2023tib}. The spatial quadrupole correlation is depicted  by the Hellings-Downs (HD) curve~\cite{Hellings:1983fr},
\e
\mu(\xi_{ab}) = \frac{1}{2} - \frac{1-\cos\xi_{ab}}{8} + \frac{3(1-\cos\xi_{ab})}{4} \ln \(\frac{1-\cos\xi_{ab}}{2}\) ,
\q
where $\xi_{ab}$ is the angular separation between a pair of pulsars $a$ and $b$.
The observation of a correlation aligned with the distinctive HD curve stands as the unequivocal "smoking gun" signature indicating successful GW detection by a PTA.

\begin{figure}[htbp]
    \centering
    \includegraphics[width=\linewidth]{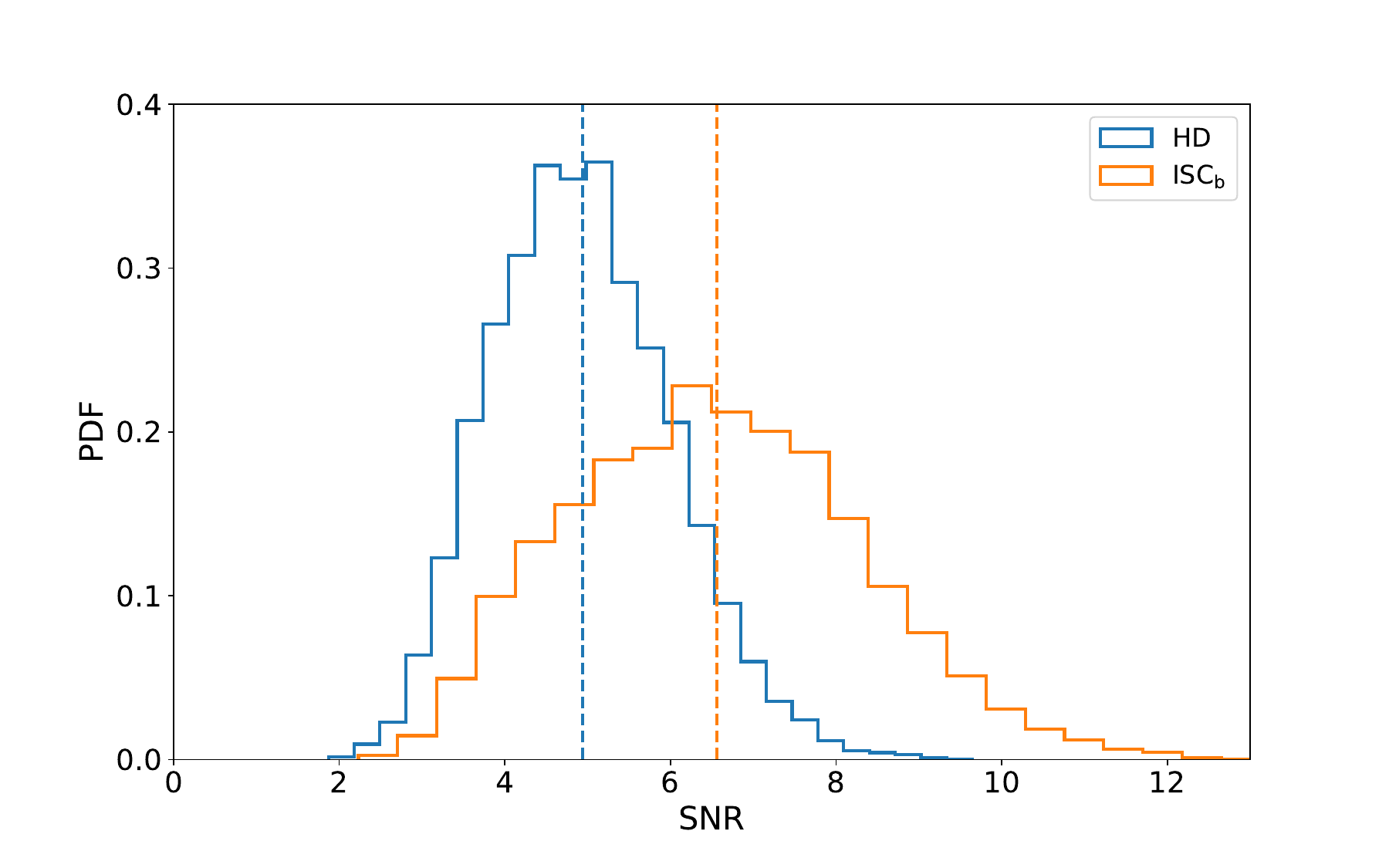}
    \caption{\label{snr_distr} \textbf{ Optimal statistic SNR distribution for different spatial correlations over the CURN noise-parameter posteriors.} The HD correlation and \text{\ISCb} ($\Gamma_{i,\text{best}}$) are analyzed individually within the Noise-Marginalized optimal statistic framework. The SNRs for the HD correlation and \text{\ISCb} are $4.9 \pm 1.1$ and $6.6 \pm 1.7$, respectively, with mean values indicated by dashed vertical lines.} 
\end{figure}

Recently, four PTA collaborations, including the North American Nanohertz Observatory for Gravitational Waves (NANOGrav)~\cite{NANOGrav:2023hde,NANOGrav:2023gor}, the European PTA~\cite{EPTA:2023sfo,EPTA:2023fyk}, Parkes PTA~\cite{Zic:2023gta,Reardon:2023gzh}, and Chinese PTA~\cite{Xu:2023wog}, have announced their respective evidence for the gravitational-wave background (GWB) to varying extents. Especially noteworthy, the NANOGrav 15-year data set demonstrates ``compelling evidence" \cite{NANOGrav:2023gor} with the Bayes factor comparing a common process with a HD correlation to that without any correlation (referred to as "common uncorrelated red noise" or CURN in the literature) ranging from $200$ to $1000$ ($212\pm 45$ in our spectrum modeling choice; the first pair of model comparisons in \Fig{BF}), and the optimal statistic signal-to-noise ratio (SNR) of $5\pm 1$ ($4.9\pm 1.1$ in our spectrum modeling choice;  blue line in the left panel of \Fig{snr_distr}). Such advancements instill great confidence that we are witnessing a pivotal moment in gravitational wave detection and revealing new phenomenons \cite{Bi:2023tib,Wu:2023hsa,Chen:2023uiz}.

\usetikzlibrary{positioning}

\begin{figure}
\centering
\begin{tikzpicture}
\tikzstyle{place}=[rounded corners,draw=black!50,fill=black!10,thick,
minimum width=1.8cm, minimum height=1.5cm,node distance=2.2cm,>=stealth,thick,align=center]
\tikzstyle{place2}=[rounded corners,draw=black!50,fill=black!10,thick,
minimum width=1.8cm, minimum height=1.5cm,node distance=2.2cm,>=stealth,thick,align=center,dashed]
\node[place][text
width=3cm] (A) {\textbf{CURN}};
\node[place] [right= of A][text
width=3cm] (B) {\textbf{HD}};
\node[place][right= of B] [text
width=3cm] (C) {\textbf{\ISCb}};
\node[place2][below=1.5cm of B] [text width=3.1cm] (D) {\textbf{HD+\Monond}};
\node[place2][below=1.5cm of C] [text width=3.1cm] (E) {\textbf{\ISCb+\Monond}};
\draw[->] (A) -- (B) node[midway, above] {\large{$\mathbf{212\pm 45}$}};
\draw[->] (B) -- (C) node[midway, above] {\large{$\mathbf{33.7\pm3.2}$}};
\draw[->,dashed] (B) -- (D) node[midway, left]
{\large{$\mathbf{2.25 \pm 0.01}$}};
\draw[->,dashed] (C) -- (E) node[midway, right]
{\large{$\mathbf{0.987\pm 0.005 }$}};
\end{tikzpicture}
\caption{\textbf{Bayes factors between models with different spatial correlations}. 
The models compared are: common uncorrelated red noise (CURN), common HD-correlated process (HD), and common correlated process for interfering sources with best-fits correlation (\text{\ISCb}; described by \Eq{Gm_i} with parameters fixed to their best-fits, $\Gamma_{i,\text{best}}(\xi_{ab})$ ). Additionally, a quasi-monochromatic monopole process in the second frequency bin (\text{\Monond}) is considered alongside the common correlated processes, i.e., HD+\text{\Monond} and \text{\ISCb+\Monond}.
All models feature variable-$\gamma$ power laws in the common spectrum. The Bayes factors, indicated along arrows, show that HD is very strongly favored over CURN,  \text{\ISCb} is strongly preferred over HD, and the presence of the additional \text{\Monond} process upon HD or  \text{\ISCb} is inconculsive, according to conventional interpretation\cite{BF}.}
\label{BF}
\end{figure}
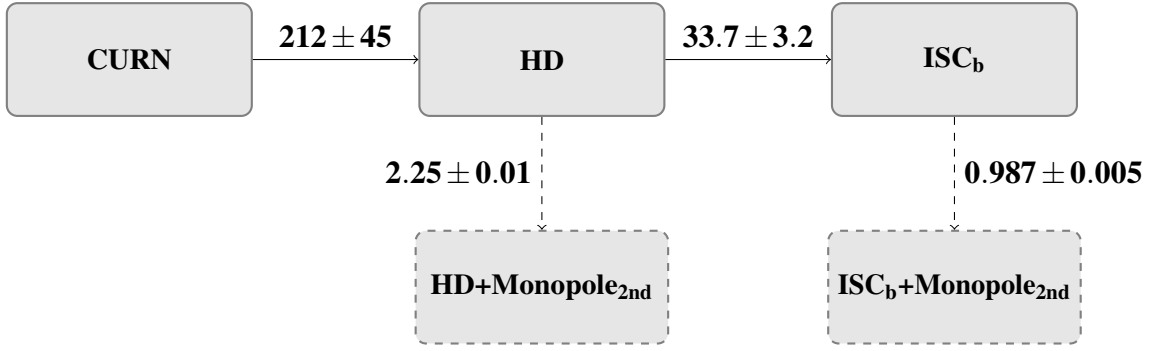

However, this is just part of the story. The HD curve serves as a prediction for a GWB originating from a collection of discrete and non-interfering sources \cite{Hellings:1983fr}. Yet, it is far more plausible that our universe hosts numerous GW sources emitting signals at frequencies close to each other~\cite{Allen:2022dzg}, on the order of $> \mathcal{O}(10^{3})$ within the frequency interval $\Delta f \sim 1/T$, with $T$ the duration of PTA experiments, typically spanning years or decades. Consequently, a realistic GWB is likely to arise from a composite of interfering sources, potentially resulting in correlations different from the HD curve.

To illustrate this scenario, we adopt the "confusion-noise" model, as outlined in~\cite{Allen:2022dzg}, which comprises a collection of sources scattered randomly across the celestial sphere. These sources emit monochromatic unpolarized GWs with random phases, all confined within the same frequency bin. The waveform for any given source is represented by
\e
h_j(t)=A_j \exp[i(2 \pi f t+\phi_j)] ,
\q
where $f$ is the frequency,  and $A_j$, $\phi_j$ are the amplitude and phase of the source, respectively. These interfering sources give rise to the spatial correlation
\begin{align}
\Gamma(\xi_{ab})&=\sum_{j,k} \mathcal{A}_j\mathcal{A}_{k} \cos(\phi_j-\phi_k)\mu(\xi_{ab},\beta_{jk}) \notag\\
&= \mu(\xi_{ab})+\sum_{j \neq k} \mathcal{A}_j\mathcal{A}_{k} \cos(\phi_j-\phi_k)\mu(\xi_{ab},\beta_{jk}) \notag\\
&:=\mu(\xi_{ab})+\Delta \Gamma(\xi_{ab}),
\label{ISC}
\end{align}
after applying ``pulsar averaging" for a large number of pairs of pulsars with the same angular separation $\xi_{ab}$ as pulsars $a$ and $b$\cite{Allen:2022dzg}.
Here, $\mathcal{A}_j\mathcal{A}_{k}=A_j A_k/\sum_{n}A_n^2$, and $\mu(\xi_{ab},\beta_{jk})$ represents the Hellings-Downs two-point function given by \Eq{mu_gm_bt}, with $\beta_{jk}$ denoting the angle between the directions to the two GW sources. When $\beta_{jk}=0$, $\mu(\xi_{ab},\beta_{jk})$ reduces to the HD curve $\mu(\xi_{ab})$. \Eq{ISC} reveals that the spatial correlations for interfering sources (referred to as ``ISC") $\Gamma(\xi_{ab})$
encapsulates contributions from two distinct components: the self-interference, which generates the conventional HD correlation, and the mutual interference among sources, which contributes to an additional term, $\Delta \Gamma(\xi_{ab})$.

Reference \cite{Allen:2022dzg} adopts a statistical perspective to understand $\Gamma(\xi_{ab})$ due to the lack of individual source information. It employs a universe ensemble comprising an infinite collection of independent universes to compute the average mean and variance of $\Gamma(\xi_{ab})$. The former precisely yields the HD curve $\mu(\xi_{ab})$, while the latter is referred to as "cosmic variance". Further details can be found in Methods section \ref{subsec1.2}. However, as we are confined to just one universe without the availability of many parallel universes for ensemble averaging, cosmic variance cannot be reduced, let alone eliminated. Consequently, it remains a fundamental limit to the precision of observing the predicted HD curve at a particular angle $\xi_{ab}$\cite{Allen:2022ksj,romano2024answers}.

Indeed, considering our Universe as a single untaken sample from the universe ensemble, we cannot predict the exact interference pattern of the sources or the resulting spatial correlations, but can only discern statistical properties such as the mean and variance of the correlations instead. However, once we isolate our Universe from the imaginary ensemble, uncertainty dissipates, and the correlations imposed by the interfering sources on the timing residuals become definite. Although we still don't know the exact form of correlation $\Gamma(\xi_{ab})$ as a function of $\xi_{ab}$ like the unequivocal HD curve $\mu(\xi_{ab})$ due to the unavailability of details of GW sources, we actually have the PTA ``ruler" to help us measure it out.

To devise a PTA-suited method for measuring the unique realization of correlations in our Universe, we resort to simulations to understand the properties of $\Gamma(\xi_{ab})$ itself. Analysis on the simulated curves shows that $\Gamma(\xi_{ab})$ can be expressed as
\e
\Gamma(\xi_{ab})=\mu(\xi_{ab})+\sum_{l=2}^{\infty} w_l P_{l}(\cos\xi_{ab}), 
\label{Gm_char}
\q
where $P_l$ and $w_l$ represent the Legendre polynomials and their corresponding coefficients at order $l$ for the mutual interference term $\Delta \Gamma(\xi_{ab})$. The distributions of $w_l$ from simulations are shown in \Fig{leg_coeffs}. The absence of contribution at $l=0$ and $l=1$ in the spatial correlation aligns entirely with the quadrupolar nature of GWs. The simulations and Legendre polynomials decomposition can refer to Methods section \ref{subsec1.3}.

We then utilize the characteristic of the novel spatial correlation shown in \Eq{Gm_char} to search for a more realistic GWB in a real PTA data set. 
The NANOGrav 15-year data set comprises observations of 68 pulsars over a period of $T=16.03$ years, with observations approximately once a month~\cite{NANOGrav:2023hde}. This extensive data set spans fluctuations distributed among roughly $\sim\mathcal{O}(10^2)$ distinct frequency bins. However, to mitigate potential correlations with excess white noise at high frequencies, the NANOGrav collaboration confines the search for the GWB (among 67 pulsars with a timing baseline exceeding 3 years) to the first 14 frequency bins~\cite{NANOGrav:2023gor}.
Notably, significant excess power for the common spectrum process is observed only in bins 1-8, with additional evidence for a quasi-monochromatic monopole process in bin 2~\cite{NANOGrav:2023gor,Agazie:2024jbf}. In this work, our analysis will specifically focus on the first 8 frequency bins for the common spectrum process, while also considering the presence of the monopole process in the second bin. For further details on the support for different correlations in individual frequency bins, refer to the analysis in Extended Data Section \ref{subsec3-1}.

GW sources within each frequency bin are expected to interfere independently, resulting in different ISCs across the frequency bins following \Eq{Gm_char}. Therefore, instead of directly searching for the conventional HD correlation $\mu(\xi_{ab})$ across all frequency bins, we measure the real correlation $\Gamma_i(\xi_{ab})$ in each frequency bin individually. As the information regarding GW sources and their resulting interfering pattern is unknown to us, we parameterize the spatial correlation in the $i$-th frequency bin $f_i=i/T$ as
\e
\Gamma_i(\xi_{ab}) = \mu(\xi_{ab}) + w_{2,i} P_2 (\cos\xi_{ab}) + w_{3,i} P_3 (\cos\xi_{ab}),
\label{Gm_i}
\q
where $i$ ranges from 1 to 8, and $w_{2,i}$ and $w_{3,i}$ respectively represent the coefficients for the Legendre polynomials of the second order $P_2 (\cos\xi_{ab})$ and the third order $P_3 (\cos\xi_{ab})$. Here, we only consider contributions from the dominant and subdominant orders $l=2$ and $l=3$, as our simulations indicate that higher-order coefficients should be smaller and can be ignored in our analysis (see \Eq{dt_w}). The coefficients $w_{2,i}$ and $w_{3,i}$ will be determined by the data set.

\begin{figure}[htbp]
    \centering
    \includegraphics[width=\linewidth]{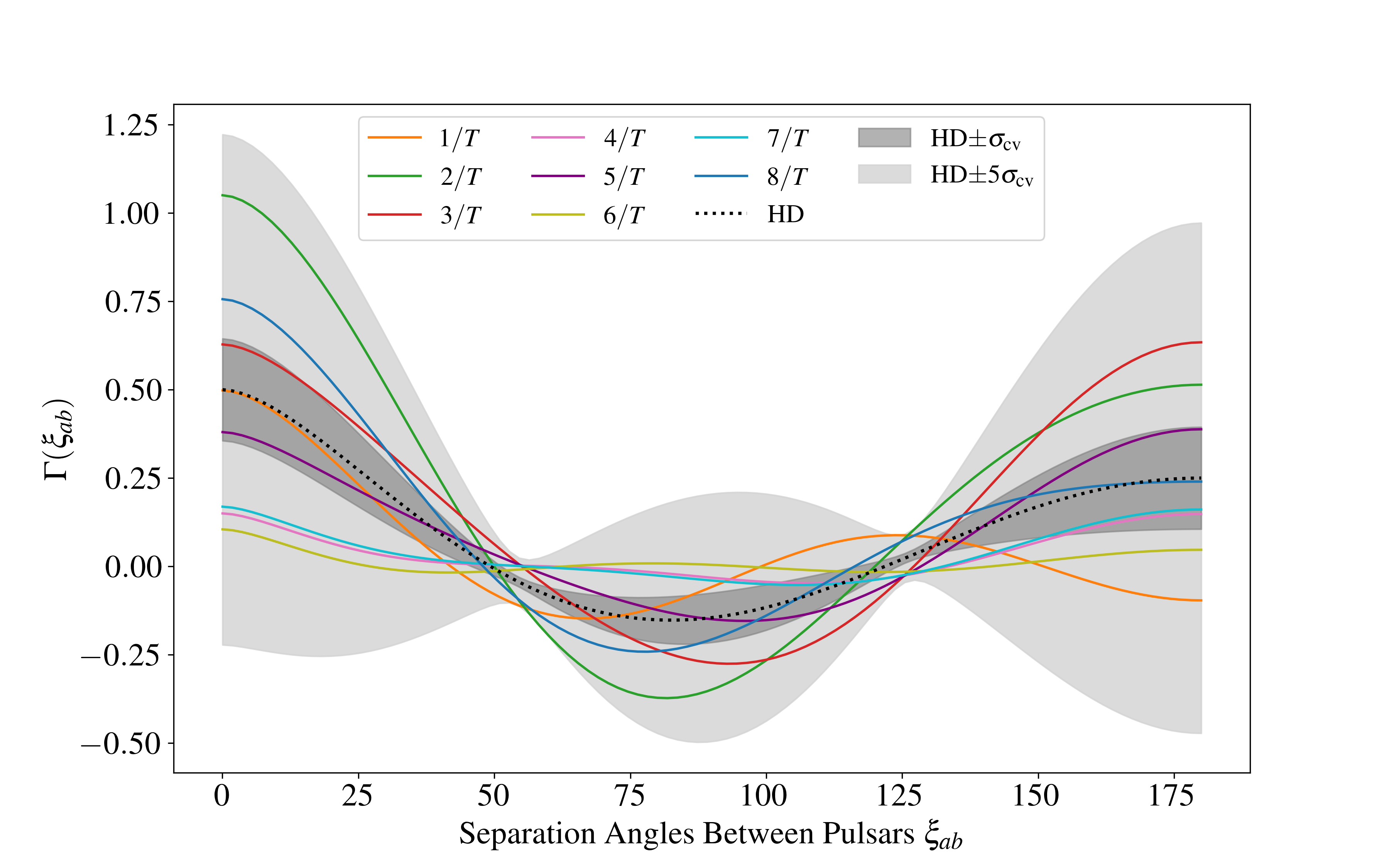}
    \caption{\label{best_fit_Gamma} \textbf{The spatial correlations for interfering sources in each frequency bin.} The colored solid lines depict the ISC curves with best-fit values, $\Gamma_{i,\text{best}}(\xi_{ab})$, with respect to the separation angles between pulsars, $\xi_{ab}$, at frequencies $f_i=i/T$, where $T$ represents the total data set time span. The statistics of the correlations within the universe ensemble, including the mean HD curve, as well as the $\sigma_{\rm cv}$ and $5\sigma_{\rm cv}$ deviation regions (where $\sigma_{\rm cv}$ denotes the root-mean-square of cosmic variance as described in Equation \Eq{CV}), respectively represented by a dotted line and dark and light grey shadows, are provided for reference. While individual ISC curves differ from the HD curve, they show consistency with cosmic variance as a whole.}
    
\end{figure} 

We utilize the Bayesian approach to infer the posteriors of the Legendre parameters $w_{2,i}$ and $w_{3,i}$ for the mutual interference term $\Delta \Gamma_i (\xi_{ab})$, as illustrated in \Fig{cv_params_dist} in Extended Data Section \ref{subsec3-2}. Although the HD correlation ($w_{2,i}=w_{3,i}=0$) is theoretically unlikely to hold in a realistic interfering-GW scenario, it remains permissible within the loosely constrained parameter posteriors due to current sensitivity limitations. However, some parameters still exhibit moderate peaks deviating from zero, suggesting that assuming the default HD correlation may lead to an underestimation of the evidence for detecting a GWB. Therefore, we fix the ISC curve $\Gamma_i (\xi_{ab})$ to their best-fit values $\Gamma_{i,\text{best}}(\xi_{ab})$ (referred to as ``\text{\ISCb}" model) in this work to optimistically assess the potential enhancement in evidence when incorporating the interference. The ISC best fits $\Gamma_{i,\text{best}}(\xi_{ab})$ in each frequency bin are shown in \Fig{best_fit_Gamma}.

Our Bayesian analysis shows a strong preference for the inclusion of 
\text{\ISCb} ($\Gamma_{i,\text{best}}$) over HD correlation within the common-spectrum signal, with the Bayes factor between the \text{\ISCb}  and HD models being $33.7\pm 3.2$. \Fig{BF} summarizes the Bayes factors between various models, comparing the CURN and HD models, and the HD and \text{\ISCb} models. 
A hybrid Bayesian-frequentist method, which employs the optimal statistic while marginalizing the pulsars' intrinsic-red-noise posterior from an initial Bayesian inference, is also employed to measure correlated excess power in PTA residuals. The distribution of the signal-to-noise ratio (SNR) for both the HD correlation and \text{\ISCb} is illustrated in \Fig{snr_distr}, with  SNRs of $4.9 \pm 1.1$ and $6.6 \pm 1.7$, respectively (means$\pm$ standard deviations). Detailed statistical analyses can be found in the Methods section \ref{sec2}.

To provide a consistent check, we also test the null hypothesis by examining the distribution of the Bayes factor (\ISCb versus CURN) and the SNR when the distinctive cross-pulsar correlations are effectively destroyed by introducing random phase shifts in the Fourier basis of the common spectrum process (the "phase shift" method ~\cite{Taylor:2016gpq}). By performing 1000 phase shifts in both Bayesian and optimal statistic analyses, we find that none of the Bayes factors exceed 1, and the highest SNR is 4.3 (only two SNRs exceed 4), consistent with the hypothesis that the data contains a GWB signal with \text{\ISCb}. However, quantitatively estimating the significance of the hypothesis or determining the false-alarm probability requires far more than 1000 phase shifts, and the immense computational resources needed for this are beyond our capability. We hope that formal PTA collaborations will address this in future data analyses, especially as enhanced sensitivity provides a more well-constrained ISC parameter space.



In conclusion, the interference among GW sources makes it unlikely to precisely recover the HD curve, despite potential advancements in telescope sensitivity, pulsar numbers, and observation duration in the future. Nevertheless, with heightened sensitivity, PTAs enable more precise measurements of the spatial correlations for interfering sources within each frequency bin, which inherits the characteristics of GW radiations while embodying the effects of interference. In this sense, our proposed strategy offers a practical pathway to detect a realistic GWB.

\section*{Methods}

\section{The spatial correlations for interfering sources and their interpretations}\label{sec1}


Pulsar timing experiments exploit the consistent arrival rates of radio pulses emitted by highly stable millisecond pulsars \cite{Detweiler:1979wn,1978SaPTA,1990ApJ...361..300F}. GWs can induce perturbations in the geodesics of these radio waves, causing redshifts in their frequencies and resulting in variations in the times of arrival (TOAs) of the pulses. These GW-induced timing residuals are expected to be spatially correlated between different pulsars \cite{Hellings:1983fr}. However, the conventionally expected HD correlation assumes discrete, non-interfering sources and thus does not reflect the realistic scenario of interfering GW sources.
In this section, we derive and interpret the spatial correlations for interfering GW sources. The derivation follows the approach outlined in \cite{Allen:2022dzg}; further details can be found in the referenced work. We adopt the natural unit system with the speed of light $c=1$, and use bold letters with hats, such as $\hat{\bm{\Omega}}$ and $\hat{\bm{p}}$, to denote unit vectors. Latin subscripts are used in various ways: ``$a$" and ``$b$" label pulsars, ``$j$" and ``$k$" label GW sources, ``$i$" labels the Fourier frequency bin, and ``$m$" and ``$n$" label the spatial components of a vector or tensor.


\subsection{The spatial correlations for interfering sources-}\label{subsec1.1}

GWs emitted by multiple unpolarized sources, each propagating towards the Earth along a $\hat{\bm{\Omega}}_{j}$ direction, induce a redshift in the frequency of radio pulses from a pulsar located at $L\hat{\bm{p}}$ relative to Earth, as described by the equation in the complex basis,
\e
Z(t)=\frac{1}{2} \sum_{j} \(F(\hat{\bm{\Omega}}_j) \Delta h_j(t, L \hat{\bm{p}}) + F^{*}(\hat{\bm{\Omega}}_j) \Delta h^{*}_j(t, L \hat{\bm{p}})\).
\q
Here, $\Delta h_j(t, L\hat{\bm{p}})$ denotes the complex GW strain difference between the Earth at time $t$ (the ``Earth term") and the pulsar at a retarded time $t-L$ (the ``Pulsar term"),
\e
\begin{aligned}
 \Delta h_j(t, L \hat{\bm{p}})&=A_j \mathrm{e}^{i\, \(2\pi f_j t+\phi_j\)}-A_j \mathrm{e}^{i\, \[2\pi f_j \(t-L(1+\hat{\bm{\Omega}}_j\cdot \hat{\bm{p}} )\)+\phi_j\]}
\end{aligned}
\q
where $A_j$, $f_j$, and $\phi_j$ represent the amplitude, frequency, and phase of the GW source, respectively; and $F(\hat{\bm{\Omega}}_j)$ symbolizes the corresponding complex antenna pattern, incorporating both the "plus" and "cross" polarization components, expressed as
\e
F(\hat{\bm{\Omega}}_j)=F^{+}(\hat{\bm{\Omega}}_j)-iF^{\times}(\hat{\bm{\Omega}}_j),
\q
with $F^{+,\times}(\hat{\bm{\Omega}}_j)$ related to the polarization tensor $\varepsilon_{mn}^{+,\times}$ as~\cite{Chamberlin:2011ev}
\e
F^{+,\times}(\hat{\bm{\Omega}}_j)=\frac{\hat{p}^{m}\hat{p}^{n}}{2(1+\hat{\bm{\Omega}}_j\cdot \hat{\bm{p}})}\varepsilon_{mn}^{+,\times}.
\q

Therefore, the time averaged correlation between pulsar $a$ and pulsar $b$ after decades of observation is 
\e
\begin{aligned}
\rho_{ab} &\equiv\overline{Z_{a}(t) Z_{b}(t)}\\
&=\sum_{j,k} \overline{c_j d_k \mathrm{e}^{2\pi i(f_j+f_k)t} \mathrm{e}^{i(\phi_j+\phi_k)} + c_j d_k^{*} \mathrm{e}^{2\pi i(f_j-f_k)t} \mathrm{e}^{i(\phi_j-\phi_k)} + c_j^{*} d_k \mathrm{e}^{-2\pi i(f_j-f_k)t} \mathrm{e}^{-i(\phi_j-\phi_k)} + c_j^{*} d_k^{*} \mathrm{e}^{-2\pi i(f_j+f_k)t} \mathrm{e}^{-i(\phi_j+\phi_k)}} \\
&= \sum_{j} \(c_j d_j^{*} + c_j^{*} d_j\) + \sum_{j\neq k}\[ c_j d_k^{*} \overline{\mathrm{e}^{2\pi i(f_j-f_k)t}} \mathrm{e}^{i(\phi_j-\phi_k)} + c_j^{*} d_k \overline{\mathrm{e}^{-2\pi i(f_j-f_k)t}} \mathrm{e}^{-i(\phi_j-\phi_k)}\] ,
\end{aligned}
\label{rhosimp1}
\q
with
\e
c_j = \frac{1}{2} A_j \[1 - \mathrm{e}^{-2\pi i f_j L_a (1 + \hat{\bm{\Omega}}_j \cdot \hat{\bm{p}}_a)}\] F_a(\hat{\bm{\Omega}}_j), \ 
d_k = \frac{1}{2} A_k \[1 - \mathrm{e}^{-2\pi i f_k L_b (1 + \hat{\bm{\Omega}}_k \cdot \hat{\bm{p}}_b)}\] F_b(\hat{\bm{\Omega}}_k).
\q
In the second line of \Eq{rhosimp1}, we nullify the time-averaging terms containing $\overline{\mathrm{e}^{2\pi i(f_j+f_k)t}}$, while the value of the remaining terms containing $\overline{\mathrm{e}^{2\pi i(f_j-f_k)t}}$ depends on the specific relationship between $f_j$ and $f_k$. When the sources emit GWs with relatively disparate  frequencies compared to the observational time scale, interference between sources occurs sparsely, causing all $\overline{\mathrm{e}^{2\pi i(f_j-f_k)t}}$ for $j\neq k$ to vanish, leaving only the self-interfering diagonal terms. Conversely, if numerous sources emit at nearly identical frequencies with $f_j\approx f_k$, non-diagonal mutual interference also persists. While PTAs exhibit sensitivity to gravitational waves across a broad frequency range of $10^{-9} \sim 10^{-7}$ Hz, each frequency bin in the Fourier expansion with a width $\Delta f \sim 1$ nHz  encompasses over $\mathcal{O}(10^3)$ sources with nearly identical frequencies. Therefore, we expect the correlation in every particular frequency bin to take the form
\e
\rho_{ab} = \sum_{j}\( c_j d_k^{*} + c_j^{*} d_k \)+ \sum_{j\neq k} \[c_j d_k^{*} \mathrm{e}^{i(\phi_j-\phi_k)} + c_j^{*} d_k \mathrm{e}^{-i(\phi_j-\phi_k)}\].
\q

The correlation discussed above involves only a single pulsar pair. However, in practice, PTAs typically monitor dozens of pulsars, leading to hundreds or even thousands of pairs. When these pairs are grouped into several angular separation bins, each bin typically contains multiple pulsar pairs with similar angular separations. Consequently, what we observe is a "pulsar averaging" correlation, which involves different pulsar pairs separated by the same angle $\xi_{ab}=\cos^{-1}(\hat{\bm{p}}_a\cdot \hat{\bm{p}}_b)$,
\e
\begin{aligned}
\langle\rho_{ab}\rangle_p =& \sum_{j, k}\[ \langle c_j d_k^{*} \rangle_p \mathrm{e}^{i(\phi_j-\phi_k)} + \langle c_j^{*} d_k \rangle_p  \mathrm{e}^{-i(\phi_j-\phi_k)}\] \\
=&\frac{1}{4}\sum_{j,k} A_j A_k \mathrm{e}^{-i(\phi_j-\phi_k)}\times\\
&\left\langle \[1\!-\!\mathrm{e}^{-2\pi i f_k L_b (1 + \hat{\bm{\Omega}}_k \cdot \hat{\bm{p}}_b)} \!-\! \mathrm{e}^{2\pi i f_j L_a (1 + \hat{\bm{\Omega}}_j \cdot \hat{\bm{p}}_a)} \!+\! \mathrm{e}^{2\pi i [f_j L_a (1 + \bm{\Omega}_j \cdot \bm{p}_a)-f_k L_b (1 + \hat{\bm{\Omega}}_k \cdot \hat{\bm{p}}_b)]} \]  F_a(\hat{\bm{\Omega}}_j)F_b^*(\hat{\bm{\Omega}}_k) \right\rangle_p + C.C.\\
=&\frac{1}{4}\sum_{j,k} A_j A_k \mathrm{e}^{-i(\phi_j-\phi_k)} \langle F_a(\hat{\bm{\Omega}}_j)F_b^*(\hat{\bm{\Omega}}_k)\rangle_p+C.C.,
\end{aligned}
\label{rhoave}
\q
where the subscript $p$ signifies the pulsar average, and $C.C.$ represent the conjugate complex term. 
The exponential functions in the square bracket stemming from the pulsar term suffer from rapid oscillation with a typical value $fL \gtrsim 10^{2}$ and thus have been eliminated \cite{Cornish:2013aba,Chamberlin:2011ev,Allen:2022dzg}. 
The pulsar average on the products of antenna pattern functions leads to 
\e
\langle\rho_{ab}\rangle_p = \frac{1}{2} \sum_{j,k} A_j A_k \cos\(\phi_j - \phi_k\) \mu(\xi_{ab}, \beta_{jk}),
\q
where $\mu(\xi_{ab}, \beta_{jk})$ denotes the HD two-point function dependent on both the pulsar pair's angular separation $\xi_{ab}$ and the angle between GW sources
$\beta_{jk}=\cos^{-1} (\hat{\bm{\Omega}}_j\cdot \hat{\bm{\Omega}}_k) $. It is specifically given by,
\e
\mu(\xi, \beta) = \left\{
	\begin{aligned}
		\frac{1}{32} &\left( 33 - 18 \cos\beta - 3 \cos^2\beta + (32 - 21\cos\beta - 6\cos^2\beta - \cos^3\beta) \cos\xi \right) \sec^4(\frac{\beta}{2}) \\
		&+ \frac{3}{2}(1 - \frac{1}{2}\cos\beta - \frac{1}{2}\cos\xi) \sec^4(\frac{\beta}{2}) \log(\sin^2(\frac{\xi}{2})), & \text{for} \ & \beta < \xi \\
		\frac{1}{16} &\left( 33 - 3 \cos\beta - (16 + 5\cos\beta + \cos^2\beta) \cos\xi \right) \sec^2(\frac{\beta}{2}) \\
		&+ \frac{3}{2}(1 - \frac{1}{2}\cos\beta - \frac{1}{2}\cos\xi) \sec^4(\frac{\beta}{2}) \log(\sin^2(\frac{\beta}{2})) , & \text{for} \ & \beta > \xi 
	\end{aligned}
	\right.
\label{mu_gm_bt}
\q
which takes the form of the HD function when $\beta=0$. Thus, we extract the spatial correlation reflecting geometric dependence as
\e
\Gamma(\xi_{ab})=\mu(\xi_{ab})+ \sum_{j \neq k} \mathcal{A}_{j}\mathcal{A}_{k} \cos\(\phi_j - \phi_k\) \mu(\xi_{ab}, \beta_{jk})=\mu(\xi_{ab})+\Delta \Gamma(\xi_{ab}),
\label{gammanorm}
\q
where $\mathcal{A}_{j}\mathcal{A}_{k}=A_j A_k/\sum_{n}A_n^2$. This expression reduces to the conventional HD correlation when only considering self-interference for discrete and non-interfering sources in previous assumptions.

Finally, we complete the consideration by examining the auto-correlation $\langle\rho_{aa}\rangle_p$, which reflects the influence of GW sources on the same pulsar. Returning to \Eq{rhoave} with $b=a$, we observe that the last exponential term in the square bracket, $\mathrm{e}^{2\pi i [f_j L_a (1 + \hat{\bm{\Omega}}_j \cdot \hat{\bm{p}}_a)-f_k L_a (1 + \hat{\bm{\Omega}}_k \cdot \hat{\bm{p}}_a)]}$, gives rise to an additional contribution $\delta_{jk}$. Thus, we obtain:
\e
\Gamma({\xi_{aa}})=2\mu(0)+\Delta\Gamma(0).
\q


\subsection{Ensemble average and variance-}\label{subsec1.2}

Ref.\cite{Allen:2022dzg} interprets the double sum term in \Eq{gammanorm}  as an uncertainty in spatial correlation arising from interference between GW sources, necessitating an ensemble of universes for its evaluation. Although $\Gamma(\xi_{ab})$ is not proportional to the HD curve $\mu(\xi_{ab})$ for any single representative universe, examining multiple representatives from the ensemble provides insight into the average and variance of $\Gamma(\xi_{ab})$, which represents the expected variation between $\Gamma(\xi_{ab})$ in any specific realization of the ensemble and the ensemble average.

One would obtain the ensemble average of $\Gamma(\xi_{ab})$ by averaging over random phases 
\e
\langle \Gamma(\xi_{ab})\rangle=\langle \Gamma(\xi_{ab})\rangle_{\phi}=\mu(\xi_{ab}),
\q
which exactly reproduces the shape of the HD curve $\mu(\xi_{ab})$.
Calculating the variance of $\Gamma(\xi_{ab})$ also involves averaging over the the angle $\beta$ between two GWs, yielding 
\e
\langle \Gamma^2(\xi_{ab})\rangle-\langle \Gamma(\xi_{ab})\rangle^2=\langle \Gamma^2(\xi_{ab})\rangle_{\phi,\beta}-\langle \Gamma(\xi_{ab})\rangle^2
=\sigma_{\rm{CV}}^2(\xi_{ab})
\q
in the limit of large number of sources with $(\sum_{n}A_n^2)^2\gg \sum_n A_n^4$ \cite{}.
$\sigma_{\rm{CV}}^2(\xi_{ab})$
is referred to as the
``cosmic variance," and its expression is given by
\e
\begin{aligned}
\sigma_{\rm{CV}}^2(\xi)
=&-\frac{15}{64}+\frac{49}{192}\cos^2\xi-\frac{3}{8}(\cos^2 \xi+3)\log\(\frac{1-\cos\xi}{2}\)\log\(\frac{1+\cos\xi}{2}\)\\
&+\frac{3}{16}(\cos \xi -1)(\cos \xi +3)\log\(\frac{1-\cos\xi}{2}\)+\frac{3}{16}(\cos \xi +1)(\cos \xi -3)\log\(\frac{1+\cos\xi}{2}\). 
\end{aligned}
\label{CV}
\q
The mean and variance on the correlation has also been effectively derived for a Gaussian ensemble on a power spectrum formalism in ~\cite{Bernardo:2022rif,Bernardo:2022xzl}.

\subsection{Specific realization in our Universe-}\label{subsec1.3}

By contrast, in our interpretation, although the exact interference pattern or resulting correlation is not predictable due to the inaccessible phase information of GW sources, it is undoubtedly definite in our unique Universe and thus promisingly measurable in an appropriate way.

To develop a method suitable for PTA to measure the unique realization of spatial correlations in our Universe, we use simulations to understand the properties of the mutual interference term $\Delta\Gamma(\xi_{ab})$.
We generate 10,000 realizations of spatial correlation for interfering sources $\Gamma(\xi_{ab})$, each comprising 1,000 GW sources uniformly distributed across the sky with random directions $\Omega_j$ and random phases $\phi_j$. Each realization yields a specific correlation curve. 

\begin{figure}[htbp]
    \centering
    \includegraphics[width=\linewidth]{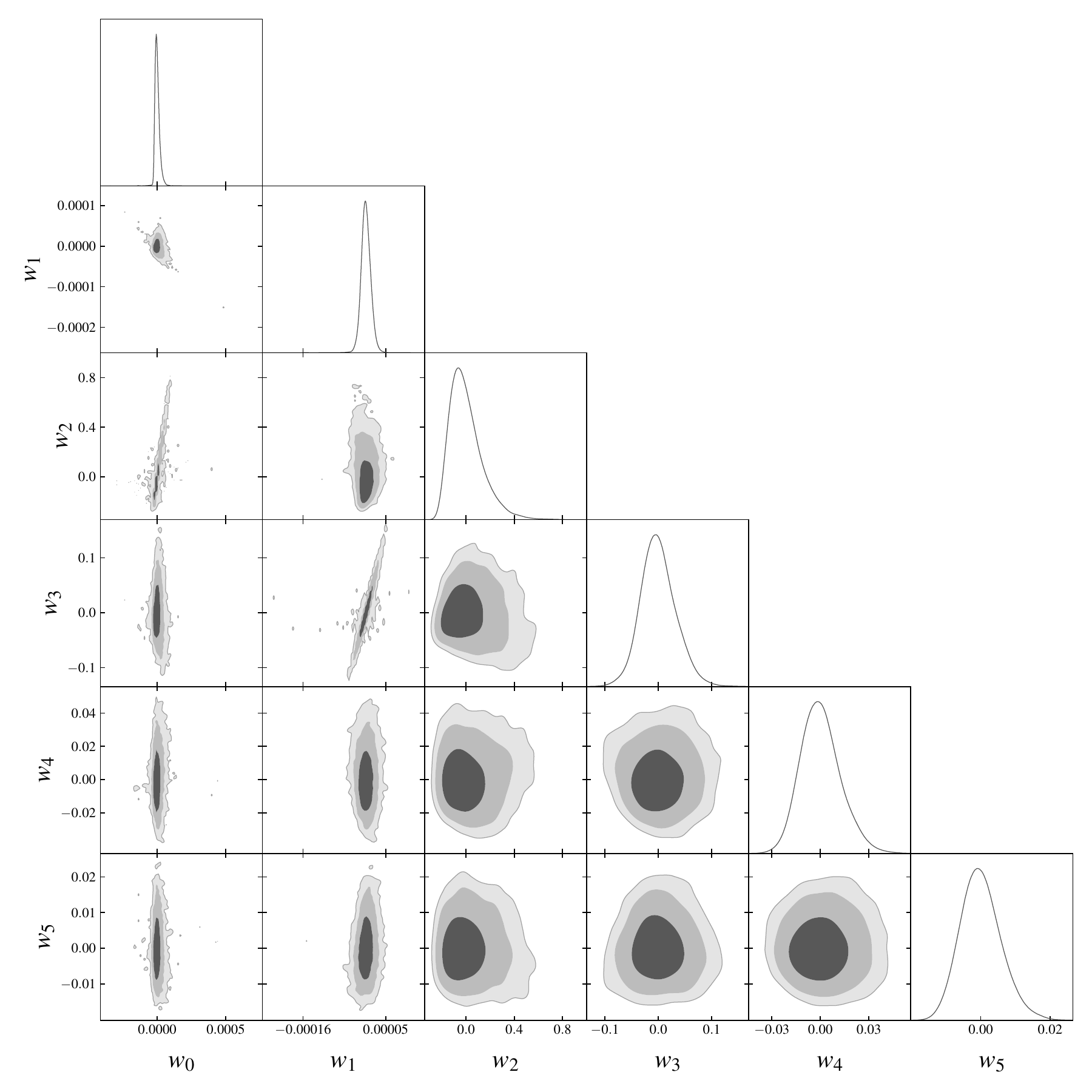}
    \caption{\label{leg_coeffs} \textbf{Distribution of Legendre coefficients for mutual interference correlation from simulations.} The coefficients $w_l$ from the Legendre polynomial decomposition \Eq{LegExp} of the interference term $\Delta\Gamma(\xi_{ab})$ in \Eq{gammanorm} are depicted in $1\sigma$, $2\sigma$ and $3\sigma$ contours up to the fifth order, based on $99.74\%$ of 10,000 simulations. Coefficients $w_0$ and $w_1$ show negligible deviations from zero, around $\mathcal{O}(10^{-5})$, indicating that $\Delta\Gamma(\xi_{ab})$ predominantly contributes at $l=2$ and higher orders. }
    
\end{figure}

In this scenario, each correlation curve, represented as a function of the pulsar angular separation $\xi_{ab}$, can be expanded using Legendre polynomials:
\e
\Gamma(\xi_{ab}) = \sum_{l=0}^{\infty} (g_l + w_l) P_{l}(\cos\xi_{ab}).
\label{LegExp}
\q
Here, $P_{l}(\cos\xi_{ab}) $ denotes the Legendre polynomials of order $l$ evaluated at $\xi_{ab}$, and $g_l$ and $w_l$ represent the Legendre coefficients of $\mu(\xi_{ab})$ and $\Delta \Gamma(\xi_{ab})$ at the corresponding order, respectively. The explicit HD curve $\mu(\xi_{ab})$ has analytical coefficients \cite{Gair:2014rwa}:
\e
g_l = \left\{
\begin{aligned}
    &0 , &\text{for} & \ l=0,1 \\
    &\frac{3}{2} (2l+1) \frac{(l-2)!}{(l+2)!} , &\text{for} & \ l \geq 2
\end{aligned}
\right.
\q
which indicates that the HD correlation has no contribution from orders $l=0$ and $l=1$. However, when this self-interference term of $\Gamma(\xi_{ab})$, i.e., $\mu(\xi_{ab})$, is extracted from simulations and subjected to the Legendre decomposition, $g_1$ and $g_2$ exhibit a tiny deviation from zero at the order of $\mathcal{O}(10^{-5})$, attributed to the precision of numerical calculation.
Comparatively, the mutual interference term $\Delta\Gamma(\xi_{ab})$ has different Legendre coefficients in each realization. The distributions of these coefficients $w_l$ up to the fifth order, based on $99.74\%$ of 10,000 realizations, are illustrated in \Fig{leg_coeffs}.

All these coefficients $w_l$ ($l=0,1,\cdots,5$) are centered around nearly zero, as expected from the term $\cos(\phi_j-\phi_k)$ in $\Delta\Gamma(\xi_{ab})$. Their root-mean-square deviations are as follows:
\e
\begin{aligned}
\Delta w_0 = 6.04\times 10^{-5},\, \Delta w_1 = 2.28\times 10^{-5}, \,\\
\Delta w_2 = 1.41\times 10^{-1}, \, \Delta w_3 = 3.38\times 10^{-2}, \,\\
\Delta w_4 = 1.24\times10^{-2},\,  \Delta w_5 = 5.92\times 10^{-3}.
\end{aligned}
\label{dt_w}
\q
We observe that the distributions of $w_0$ and $w_1$ also exhibit a negligible deviation from zero, on the order of $\mathcal{O}(10^{-5})$. Therefore, it is convincing to conclude that the $\Delta\Gamma(\xi_{ab})$ term also contributes negligibly at $l=0$ and $l=1$. In other words, the entire spatial correlation for interfering sources $\Gamma(\xi_{ab})$ predominantly contributes at $l=2$ and higher orders, with root-mean-square deviations decreasing for higher $l$, reflecting the quadrupole nature inherent in GW radiation. 
Furthermore, it is also worth noting that the root-mean-square deviations of the coefficients $\Delta w_2$ to $\Delta w_5$
based on our simulations are in  agreement with the prediction from Ref. \cite{Allen:2024uqs}, given by
\e
\Delta w_i=\frac{1}{\sqrt{N_f}} \frac{3}{2}\frac{\sqrt{(2l+1)}}{(l+2)(l+1)l(l-1)}
\q
after including the factor $3/2$ for consistent normalization, when combining $N_f$ observational frequency bins in which the GW signal dominates the noise.

\section{Main analysis}\label{sec2}

Our goal is to extract the GWB signal from the timing residuals and provide an appropriate interpretation. This requires accurate modeling of both the noise and signal in the data set, as well as proper statistical analysis, including parameter estimation, candidate model selection, and assessing the signal's relative strength. This section presents the data model and statistical analysis methods used in the main text.

\subsection{Model-}\label{subsec2.1}

Several factors, including timing model inaccuracies, white measurement noise, and pulsar intrinsic red noise, can also cause timing residuals in TOAs \cite{NANOGrav:2023gor,NANOGrav:2023ctt}. Therefore, these factors must be modeled alongside the putative signal when searching for the GWB. 


The timing-uncertainties is modeled by $M\bep$, with $M$ the design matrix basis and $\bep$ an offset vector of timing model parameters \cite{Chamberlin:2014ria,NANOGrav:2015aud}. White noise is characterized by three parameters:"EFAC" scales TOA uncertainties, "EQUAD" adds variance to TOA uncertainties in quadrature, and "ECORR" represents per-epoch variance common to all sub-bands \cite{Ellis:2014xgh,NANOGrav:2015qfw,NANOGrav:2015aud,NANOGrav:2023ctt}. Red noise is typically modeled using spectrum components $\Phi_{\rm{RN}, i}$ of a power-law form under the Fourier basis, where each pulsar has one amplitude parameter $A_{\rm{RN}}$ and one index parameter $\gamma_{\rm{RN}}$ \cite{Shannon:2010bv}. The common red signal, such as a GWB, is also commonly assumed to take power-law spectrum components but additionally incorporates spatial structure among pulsars encoded by $\Gamma_{i}(\xi_{ab})$,
\e
\Phi_{ab,i}=\Gamma_{i}(\xi_{ab})\Phi_{{\rm{gw}},i}=\Gamma_{i}(\xi_{ab})\frac{A_{\rm{gw}}^2}{12\pi^2}\frac{1}{T}\(\frac{f_i}{f_{\yr}}\)^{-\gm_{\rm{gw}}}f_{\yr}^{-3}.
\label{spect}
\q
Here, $f_i=i/T$ with $T=16.03$ yr, $A_{\rm{gw}}$ denotes the amplitude of the GW spectrum measured at the reference frequency $f_{\yr}=1/\text{yr}$, and $\gamma_{gw}$ is the power index. We use 30 frequency modes ($i \in [1,30]$) for the red noise modeling and 8 ($i \in [1,8]$) for the common signal \cite{NANOGrav:2023gor,Agazie:2024jbf}.

\subsection{Bayesian analysis-}\label{subsec2.2}

The Bayesian methodology is commonly used in GWB searches within PTA data set for parameter estimation and model selection.

Assuming stochastic processes in timing residuals $\dt \bbt $ are Gaussian and stationary, the PTA likelihood is evaluated using a multivariate Gaussian function \cite{Ellis:2014xgh},
\e
L(\dt \bbt|\Theta,\bep)=\frac{1}{\sqrt{(2\pi) {\rm{det}}(C)}}\exp\(-\frac{1}{2}\mb{r}^{\rm{T}}C^{-1}\mb{r}\).
\label{Likelihood}
\q
Here, $\mathbf{r} = \dt \mathbf{t} - M \bep$ accounts for the contribution from stochastic noise and signals by subtracting the timing model uncertainty $M \bep$ from the residuals. The design matrix $M$ is determined using the timing software \texttt{TEMPO2} \cite{Hobbs:2006cd,Edwards:2006zg}, and the timing model uncertainty parameters $\bep$ with uniform priors can be marginalized to obtain the likelihood $L(\dt \bbt|\Theta)$. 
The total covariance matrix $C = \langle\mathbf{r}\mathbf{r}^{\text{T}}\rangle = C_{\text{noise}} + C_{\text{signal}}$ is the sum of block diagonal covariance matrices for the noise of individual pulsars and the covariance of the correlated common red signal among all pulsars, described by a set of parameters $\Theta$.

From Bayes's theorem, the posterior probability is
\e
P(\Theta|\dt \bbt)=\frac{L(\dt \bbt|\Theta)\pi(\Theta)}{Z},
\q
where $\pi(\Theta)$ is the prior and $Z$ is the Bayesian evidence, given by the integral of the likelihood multiplied by the prior over the prior volume,
\e
Z=\int L(\dt \bbt|\Theta)\pi({\Theta}) d{\Theta}.
\q
For a specific parameter $\theta$, its estimation is obtained by marginalizing the posterior probability over other unwanted parameters,
\e
P(\theta|\dt \bbt)=\int P(\Theta|\dt \bbt) d{\Theta'},
\q
where $\Theta'$ includes all parameters in $\Theta$ except $\theta$.

In our search for a GWB, we fix the white noise parameters at their maximum likelihood values obtained from the single pulsar noise analysis and simultaneously assess the parameters of red noise and the common signal \cite{EPTA:2015qep}, with priors given by:
\e
\begin{aligned}
\log_{10}A_{\rm{RN}}&\sim \rm{Uniform}(-20,-11),\qquad\gm_{\rm{RN}}\sim \rm{Uniform}(0,7),\\
\log_{10}A_{\rm{gw}}&\sim \rm{Uniform}(-18,-11), \qquad
\gm_{\rm{gw}}\sim \rm{Uniform}(0,7),\\
w_2 &\sim {\rm{Uniform}}(-0.7,0.7),\,\, \qquad w_3 \sim \rm{Uniform}(-0.2,0.2).
\end{aligned}
\label{priors}
\q
It should be noted that the priors for the Legendre parameters $w_2$ and $w_3$ are set to approximately five times their root-mean-square deviations as given in \Eq{dt_w}.

Model selection between two hypothesises, $H_1$ and $H_0$, relies on the ratio of their corresponding Bayesian evidence, known as Bayes factor
\e
\rm{BF}=\frac{Z_{1}}{Z_{0}}.
\q
Typically, $\rm{BF}\in(1,3)$, $\rm{BF}\in(3,20)$, $\rm{BF}\in(20,150)$  and $\rm{BF}>150$ are interpreted as ``not worth more than a bare mention", positive, strong and very strong evidence for $H_1$ against $H_0$, respectively \cite{BF}. 

In data analysis, we utilize the \texttt{Enterprise} package \cite{enterprise} to calculate the likelihood and \texttt{PTMCMCSampler} package \cite{PTMCMCSampler} for posterior evaluation. Model selection is facilitated by the product-space sampling method, which introduces a hyperparameter toggling between the likelihoods of the two models. The Bayes factor is determined by the ratio of the sample fractions using one model to that using the other model. \cite{10.2307/2346151,10.2307/1391010,Hee:2015eba,Taylor:2020zpk}

\subsection{Optimal statistic analysis-}\label{subsec2.3}

The optimal statistic is designed to measure correlated excess power in PTA residuals~\cite{Anholm:2008wy,Chamberlin:2014ria,Vigeland:2018ipb,NANOGrav:2023gor,Pol:2022sjn}.
The optimal estimator of the signal amplitude $A_{\rm{gw}}$ can be defined via a log-likelihood ratio between the model with both spatially correlated GWB and uncorrelated noise and the model with only noise components and is given by \cite{Chamberlin:2014ria},
\e
\hat{A}^2 = \frac{\sum_{ab} \mb{r_a}^{\rm{T}} \mb{P}_a^{-1} \tilde{\mb{\Phi}}_{ab} \mb{P}_b^{-1} \mb{r_b}} {\sum_{ab} {\rm tr} \[\mb{P}_a^{-1} \tilde{\mb{\Phi}}_{ab} \mb{P}_b^{-1} \tilde{\mb{\Phi}}_{ba}\]} ,
\q
where $\mb{P}_a=\langle \mb{r_a} \mb{r_a}^{\rm{T}} \rangle$ represents diagonal terms of covariance matrix, and $\tilde{\mb{\Phi}}_{ab}$ denotes amplitude-independent off-diagonal terms, $ \tilde{\mb{\Phi}}_{ab}=\langle \mb{r_a} \mb{r_b}^{\rm{T}} \rangle /A_{\rm{gw}}^2$. In the absence of a cross-correlated signal, the expectation value $\langle\hat{A}^2\rangle=A_{\rm{gw}}^2$ vanishes and its standard deviation is given by
\e
\sigma_{\hat{A}^2} = \(\sum_{ab} {\rm tr} \[\mb{P}_a^{-1} \tilde{\mb{\Phi}}_{ab} \mb{P}_b^{-1} \tilde{\mb{\Phi}}_{ba}\]\)^{-1/2} .
\q
Thus, we can define the signal-to-noise ratio (SNR) for the power in the cross-correlations by dividing the observed $\hat{A}^2$ by its standard deviation as follows \cite{NANOGrav:2023gor}:
\e
\rm{SNR} = \frac{\hat{A}^2}{\sigma_{\hat{A}^2}}.
\q

\setcounter{section}{0}
\renewcommand\thesection{\arabic{section}}

\section*{Extended Data Section}\phantomsection\label{sec3}


\section{Correlations in individual frequency bin}\phantomsection\label{subsec3-1}

In Methods section 1.2, we have demonstrated that the inter-pulsar correlation induced by GWs, even when accounting for interfering sources, begins contributing from the $l=2$ order in the Legendre polynomials decomposition. This implies that the GW signal should exhibit no power at $l=0$ and $l=1$.
However, sources other than GWs can generate spatially correlated power at these two orders. For instance, clock errors in terrestrial time standards result in an identical  delay among all pulsars, thereby creating monopolar correlations at $l=0$, while errors in solar-system ephemerides lead to Roemer delays with dipolar correlations at $l=1$. To avoid possible excess power at $l=0$ and $l=1$ leaking to higher orders and contaminating the measurement of $\Gamma_{i}(\xi_{ab})$ (\Eq{Gm_i}), we examine each individual frequency bin with the ``Monopole+Dipole+ISC" model with free spectrum. This model allows for the simultaneous search for the three different correlated signals, with the Fourier components $\Phi_{\text{mono},i}$, $\Phi_{\text{dip},i}$, and $\Phi_{\text{gw},i}$ varying independently in each frequency bin.

\begin{figure}[htbp]
    \centering
    \includegraphics[width=\linewidth]{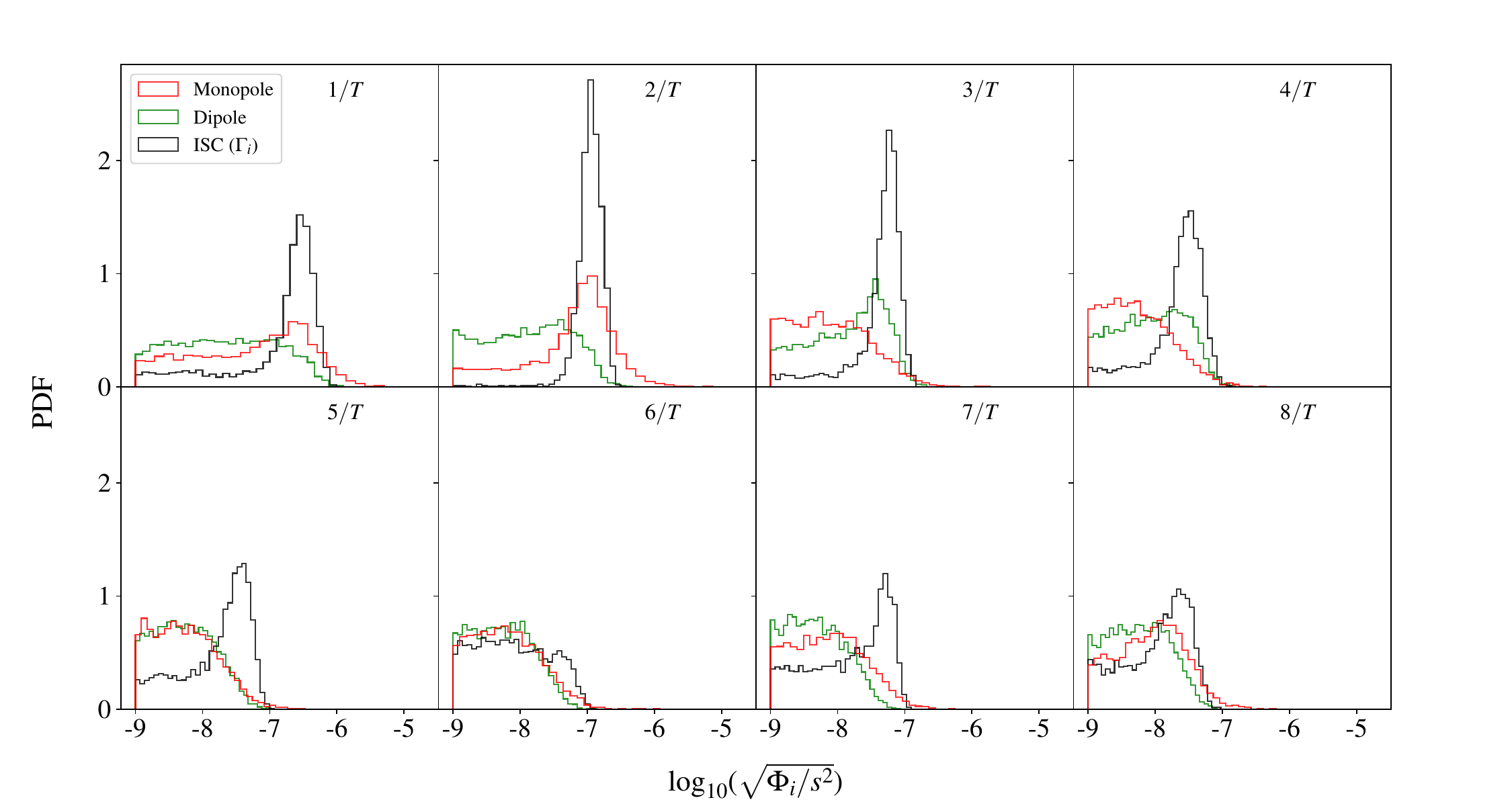}
    \caption{\label{3comps_spectrum} \textbf{Posteriors of the Fourier component variance for different correlations in individual frequency bin.} The red, blue, and black lines represent the posteriors of $\Phi_{\text{mono},i}$, $\Phi_{\text{dip},i}$, and $\Phi_{\text{gw},i}$, which are the independent Fourier component variances for the monopole correlation at Legendre order $l=0$, dipole correlation at $l=1$, and spatial correlation for interfering sources at $l=2$ and higher orders, respectively, at frequencies $f_i = i/T$, where $T$ is the 16.03-year extent of the data set. These posteriors are obtained when all correlations are simultaneously searched. Excess interfering-correlated power is observed in bins 1-8 (weak in bin 6), excess monopole-correlated process is only observed in frequency bin 2, and no significant excess dipole-correlated power is observed in any bins. These results align with those reported in \cite{NANOGrav:2023gor}.}
\end{figure} 

The posteriors of the Fourier component variance $\Phi_i$ for each correlated signal in the joint model, as depicted in \Fig{3comps_spectrum}, reveal evidence of a monopole process at frequency bin 2, while no obvious excess monopole or dipole power is observed in any other bins.

\section{The spatial correlations for interfering sources in individual frequency bin}\phantomsection\label{subsec3-2}

\begin{figure}[htbp]
    \centering
    \includegraphics[width=0.8\linewidth]{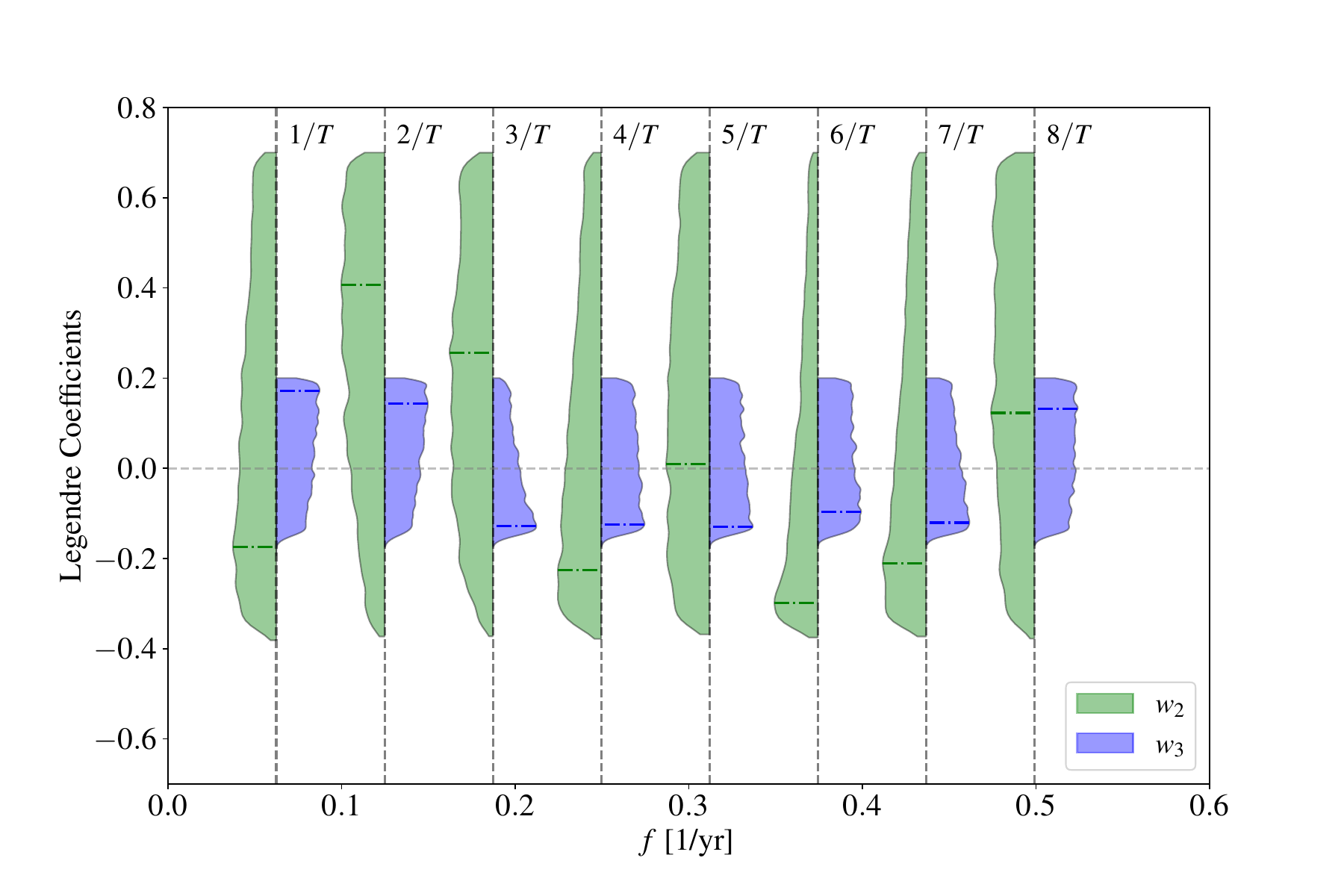}
    \caption{\label{cv_params_dist} \textbf{Posteriors of the Legendre parameters in the spatial correlations for interfering sources.} The half violins in green and purple represent the posterior distributions of the parameters $w_{2,i}$ and $w_{3,i}$ of the Legendre polynomials at the second and third orders, respectively, for the parameterized ISC in \Eq{Gm_i} at frequencies $f_i=i/T$, where $T$ denotes the extent of the data set. Dotted dashed-lines indicate the best-fit values for corresponding parameters.}
\end{figure}

In searching for the spatially correlated GWB signal described by \Eq{spect}, with a power-law spectrum and a parameterized ORF given in \Eq{Gm_i}, and priors specified in \Eq{priors}, we obtain the posterior distributions for the parameters in the parameterized ORF, as shown in \Fig{cv_params_dist}.

\section*{Acknowledgements}
This work is supported by the grants from NSFC (Grant No.~12250010, 11991052), Key Research Program of Frontier Sciences, CAS, Grant No.~ZDBS-LY-7009. We acknowledge the use of HPC Cluster of ITP-CAS. 

\section*{Author contributions statement}

QGH was the primary driver of the project. YMW proposed simulations to explore different realizations of spatial correlations across frequency bins. QGH and YCB conceptualized the Legendre decomposition approach for analyzing the simulated results. YMW conducted Bayesian analysis, while YCB performed optimal statistic analysis. The initial manuscript was primarily authored by YMW, with substantial contributions from YCB in the methods section, all under the instructive guidance of QGH.  All authors participated in interpreting and discussing the results, as well as in commenting on and editing the text.

\section*{Data Availability}
The data used for the analyses presented is publically available from \href{https://data.nanograv.org}{https://data.nanograv.org}.

\section*{Code Availability}
The code used for the analyses is available from the authors upon reasonable requests. \\

\section*{Correspondence}
Correspondence and requests for materials should be addressed to YMW (wuyumei@yzu.edu.cn), YCB (biyanchen@itp.ac.cn) and QGH (huangqg@itp.ac.cn). 

\section*{Competing Interests Statement}

The authors declare no competing interests.


\end{document}